\documentclass[aps,prl,twocolumn]{revtex4-2}

\usepackage{amsmath}
\usepackage{epsfig}
\usepackage{amssymb}
\usepackage{graphicx}

\begin{document}

\title{Parity as the foundation of the non-relativistic spin-statistics connection}
\author{Dmitri V. Averin}
\affiliation{Department of Physics and Astronomy, Stony Brook University, SUNY,
Stony Brook, NY 11794-3800}

\begin{abstract}

It is shown that the symmetry under parity of the wavefunctions of two identical particles with an arbitrary spin $s$ in three spatial dimensions accounts for the appropriate wavefunction exchange statistics under the permutations of particles. The standard properties of the angular momentum in non-relativistic quantum mechanics account for the sign factor $(-1)^{2s}$ that the wavefunctions acquire under the permutation of coordinates of the two particles, without any additional requirements, directly relating spin and the particle exchange statistics in the non-relativistic context.

\end{abstract}

\maketitle

\begin{figure}[t]
\centering
\includegraphics[width=0.3\textwidth]{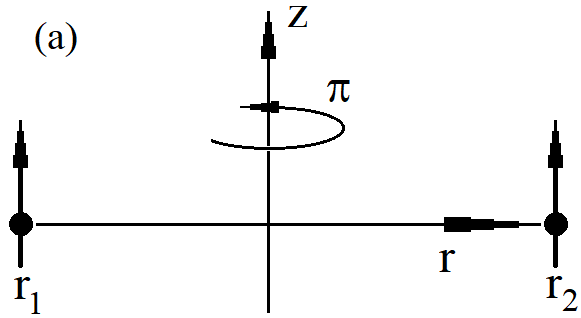}
\includegraphics[width=0.3\textwidth]{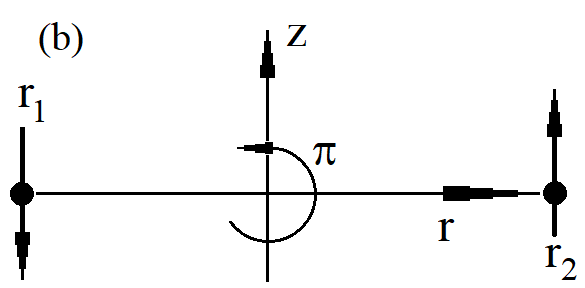}
\caption{{\protect\footnotesize {Diagram of the rotations by angle $\pi$ that interchange two particles with (a) the same and (b) opposite spin orientations. } }}
\label{fig1} \end{figure}

One of the most basic results of the theory of angular momentum in non-relativistic quantum mechanics is that the $2\pi$-rotation of a particle with a half-integer spin is not an identity transformation but reverses the sign of the particle wavefunction. This result is well established in theory -- see, e.g., \cite{aharonov,bernstein}, and was observed in experiments with neutron interferometry \cite{rauch,werner}. Such a change of sign upon $2\pi$-rotation is absent for particles with integer spin. Since the permutation of two particles can be implemented as rotation (for illustration, see Fig.~\ref{fig1}), it is natural to try to connect different rotation properties of the half-integer and integer-spin particles to the difference in their exchange statistics: the fermionic change of sign of the wavefunctions upon the permutation of coordinates of the half-integer-spin particles versus bosonic property of the wavefunctions symmetric with respect to the permutation of coordinates of integer-spin particles. Over the years, there have been attempts to elevate this semi-quantitative connection between the spin rotation properties and particle exchange statistics into a consistent proof of the spin-statistics connection completely within the context of the non-relativistic quantum mechanics (see \cite{ss1,ss2,ss3,ss4,ss4c,css4,ss5,ss6} and references therein) without the more standard recourse to quantum field theory. Although these attempts gained some acceptance even in quantum field theory textbooks \cite{schwartz}, they also encountered considerable criticism \cite{css1,css2,css3}, and still have some intrinsic unresolved issues \cite{ss4,css4}.

To discuss this more precisely, it should be noted that the non-relativistic spin-statistics connection is established by implementing the operation $E$ of the permutation of the two identical particles as a transformation of the appropriately chosen coordinate system through appropriate rotation. The rotation is chosen so that it interchanges the particles and brings the two-particle system as a whole to the same state. To achieve this, the origin of the coordinate system is taken at the center-of-mass of the system, and the $z$ axis -- orthogonal to $\vec{r}=\vec{r}_2-\vec{r}_1$, where $\vec{r}_1$ and $\vec{r}_2$ are the position vectors of the two particles (Fig.~1). The general rotation operator $R$ can be expressed then (see, e.g., \cite{sakurai}) as $R=e^{-i\hat{n}\vec{J} \theta}$ through the operator of the total angular momentum $\vec{J}$ of the two particles, $\vec{J}=\vec{L}+\vec{S}_1 +\vec{S}_2$, where $\vec{L}$ is the orbital momentum of the particles relative to the origin of the coordinate system, while $\theta$ is the angle, and $\hat{n}$ -- the unit vector along the axis of the rotation. For particles with non-vanishing spins $\vec{S}_{1,2}$, the required rotation depends on the spin state of the particles. Quantitatively, if the magnitude of the spins is $s$, an arbitrary two-particle state $|\psi \rangle$ can be expressed as
\begin{equation} | \psi \rangle= \sum_{m_1,m_2}\psi_{m_1,m_2} (\vec{r}_1,\vec{r}_2) |m_1,m_2\rangle \, ,\label{wf1} \end{equation}
where $m_1$ and $m_2$ are the $z$-components of the spins in the chosen coordinate system:  $m_{1,2}=-s,-s+1, \, ...\, s$.

If the particles are in the same spin state, $m_2=m_1\equiv m$, the $\pi$ rotation around the $z$ axis \cite{ss1,ss5,ss6,schwartz} interchanges them and returns the system to the same state (Fig.~1a). Indeed, all three contributions to the total angular momentum $\vec{J}$ commute among themselves, and their effects in the rotation operator $R$ can be considered individually. If the particle positions $\vec{r}_{1,2}$ lie in the $x-y$ plane, the orbital part of $R$ interchanges them: $e^{-i\pi L_z}\psi(\vec{r}_1,\vec{r}_2)=
\psi(\vec{r}_1,\vec{r}_2)$. To avoid including any dynamic phase in this relation, one assumes that the particles do not have any orbital angular momentum, $l=0$. Then
\[e^{-i\pi J_z}\psi_{m,m} (\vec{r}_1,\vec{r}_2)=e^{-i2\pi m} \psi_{m,m} (\vec{r}_2,\vec{r}_1) \]
\[=(-1)^{2s} \psi_{m,m} (\vec{r}_2,\vec{r}_1) \, , \]
where the last equality takes into account that $m$ can differ from $s$ only by an integer. The factor $(-1)^{2s}$ in this expression provides the sought connection between the particle spin and their exchange statistics.

If the particles are in the ``opposite'' spin states, $m_2=-m_1\equiv m$, the required rotation (which returns the system to the same state) is again by angle $\pi$, but around the $y$ axis \cite{ss3,ss6}, chosen to be orthogonal to $\vec{r}$ (Fig.~1b). Under the assumptions concerning the orbital part of the wavefunction similar to those above, one has
\[e^{-i\pi J_y}\psi_{-m,m} (\vec{r}_1,\vec{r}_2)= (-1)^{2s} \psi_{m,-m} (\vec{r}_2,\vec{r}_1) \, . \]
This equation follows directly from the appropriate matrix elements of the Wigner's rotation matrices \cite{sakurai}: $d^{(s)}_{-m,m}(\pi) =(-1)^{(s-m)}$.

Thus, the permutation of the two identical particle implemented as a rotation produces the sign consistent with the spin-statistics relations for the two types of the particle spin configurations. It is usually argued from this that it is only a technical issue to extend this logic to an arbitrary quantum state of the particles. This assertion, however, does not seem plausible, since the required rotation is state-dependent, and it is not even evident that there is an appropriate rotation with desired properties for any spin state. For instance, one can consider a state that exists for any $s$ and corresponds to the total angular momentum zero of the two spins. This state is a scalar and will not change under any rotation, regardless of the value of $s$, clearly making it impossible to reproduce by rotation only the sign required by the spin-statistics relation for half-integer $s$. From this perspective, it is not unexpected that a rigourous implementation of the particle permutation through rotation for a general spin state produces meaningless results, as demonstrated explicitly in Appendix D of \cite{ss4}.

The goal of this work is to suggest that the parity transformation (as opposed to rotations) realizes the particle exchange in a way that immediately establishes the spin-statistics connection. Before showing this explicitly, it is useful to argue why the parity is an appropriate tool for this purpose. The main reason is that the parity transformation interchanges the two particles in an arbitrary spatial  configurations. (For rotations, the particles are interchanged only if their orbital wavefunction is confined to the plane orthogonal to the rotation axis.) If one introduces the center-of-mass and the relative coordinates  $\vec{R}=(\vec{r}_1+\vec{r}_2) /2$ and $\vec{r}=\vec{r}_2 -\vec{r}_1$ for two identical particles, it is seen immediately that in the coordinate system with the origin at the center of mass $\vec{R}$ (as in Fig.~1), the parity transformation $P$: $\vec{r}\rightarrow -\vec{r}$ interchanges the two coordinate vectors $\vec{r}_1 \leftrightarrow \vec{r}_2$. This means that the action of $P$ on an arbitrary orbital part of the two-particle wavefunctions (excluding, as above, the dynamic phase) is simply
\begin{equation}
P\psi (\vec{r}_1,\vec{r}_2) =\psi (\vec{r}_2,\vec{r}_1) \, .
\label{par1}  \end{equation}
An additional useful property of the parity transformation in the present context is that it appropriately distinguishes the case of the 3-dimensional space from the 2-dimensional one, where the parity transformation separate from rotations does not exist. This is in agreement with the fact that the 2D particles can have anyonic exchange statistics that is different from the bosonic or fermionic statistics. Finally, it should be mentioned that while there is no simple connection between the parity-transformation properties of the individual particles and their spins (because the understanding of the intrinsic parity of spinors is quite limited -- see, e.g., \cite{ll4}), as shown below, one can still establish a direct and simple connection between the parity properties of identical particles and their spin values in a way that avoids the issue of spinor parity.

To understand the particle exchange implemented as the parity transformation quantitatively, one needs to start with the states that are transformed into themselves by this operation, i.e., the states for which the transformation returns both the orbital and spin configuration of the two particles to its initial form.  Since the parity conserves angular momentum, these are the states with the well-defined total momentum of the two particles. and one needs to consider the general two-particle state $|\psi \rangle$ not in the basis $|m_1,m_2\rangle$ (\ref{wf1}) of the two spins, but in the basis of the total angular momentum  $\vec{j}=\vec{s}_1+\vec{s}_2$:
\begin{equation} | \psi \rangle= \sum_{j,m}\chi_{j,m} (\vec{r}_1,\vec{r}_2) |j,m\rangle \, ,\label{wf2} \end{equation}
where $j$ is the magnitude of $\vec{j}$, which ranges from $0$ to $2s$, and $m$ is the $z$-component of $\vec{j}$.
Transition between the two bases, (\ref{wf1}) and (\ref{wf2}), is enacted by the standard ``Clebsch-Gordon'' coefficients (CGCs) $C(j,m;m_1,m_2)=\langle j,m|m_1,m_2\rangle$:
\begin{equation} \chi_{j,m} (\vec{r}_1,\vec{r}_2) =\sum_{m_1,m_2} C(j,m;m_1,m_2) \psi_{m_1,m_2} (\vec{r}_1,\vec{r}_2)\, , \label{CGS1} \end{equation}
where the sum is taken over all values of $m_1,\, m_2$ consistent with the conditions $m_1+m_2=m$ and $m_{1,2}\in [-s,s]$.
CGCs are all real, and since the transformation (\ref{CGS1}) is a unitary transition between the two bases, the inverse transformation has the same coefficients $C(j,m;m_1,m_2)$:
\begin{equation} \psi_{m_1,m_2} (\vec{r}_1,\vec{r}_2) =\sum_j C(j,m;m_1,m_2) \chi_{j,m} (\vec{r}_1,\vec{r}_2)\, , \label{CGS2} \end{equation}
where, again, $m=m_1+m_2$, but the sum is now taken over all values of $j$ in the range consistent with the values of $m$ and $s$: $|m|\leq j \leq 2s$.

The action of the parity transformation $P$ on the states with the total angular momentum $j$ of the two spins is
\begin{equation} P\chi_{j,m} (\vec{r}_1,\vec{r}_2) =(-1)^{j}\chi_{j,m} (\vec{r}_2,\vec{r}_1) \, . \label{par2} \end{equation}
This relation is usually derived for orbital angular momentum from the properties of the spherical harmonics (see, e.g., \cite{sakurai}),  but for integer $j$ relevant in this discussion: $j\in [0,2s]$, this derivation can be provided regardless of the physical origin of the angular momentum. Note also that for the same reason integer $j$, Eq.~(\ref{par2}) is independent of the uncertainty in assignment of the intrinsic parity to spinors mentioned above.

While the parity transformation in the center-of-mass system interchanges the orbital coordinates of the two identical particles [as in Eq.~(\ref{par2})], it does not interchange their spin variables. A somewhat subtle reason for this is that, in the non-relativistic context, a particle spin, as the angular momentum produced by the motion with no total linear momentum, can not be assigned to any particular point in space: it is the same relative to all spatial points. No transformation of the coordinate reference frame can transfer the spin operators. This means that complete permutation $E$ of the coordinates of the particles with non-vanishing spins, requires, besides the parity transformation (\ref{par2}), an explicit permutation $E_s$ of the spin variables: 
\begin{equation} E=E_sP \, \label{perm} \end{equation}
with $E_s$ defined by $E_s |m_1,m_2\rangle =|m_2,m_1\rangle$. Thus, to find the transformation properties of the wavefunctions $\chi_{j,m} (\vec{r}_1,\vec{r}_2)$ under the particle permutaion $E$, one needs to find the symmetry of the CGCs $C(j,m;m_1,m_2)$ with respect to the spin permutation $E_s$.

This can be done in two steps. First, considering the usual raising/lowering operators $j^{(\pm)}$ for momentum $\vec{j}$:
\begin{equation} j^{(\pm)}|j,m\rangle= [(j\mp m)(j\pm m+1)]^{1/2}|j,m\pm 1\rangle \label{e1} \end{equation}
one notes that these operators are symmetric with respect of the permutation of the two spins: $j^{(\pm)} = s^{(\pm)}_1 + s^{(\pm)}_2$. This means that all $2j+1$ states with the same $j$ and different $m\in [-j,j]$ have the same symmetry properties that coincide, e.g., with those of the state $|j,j\rangle$ with $m=j$. Then, the symmetry of the state $|j,j\rangle$ can be determined from the recurrence relations for these states that follow from the equation they satisfy:
\begin{equation}  j^+|j,j\rangle=(s^+_1 + s^+_2)|j,j\rangle=0\, .\label{e2} \end{equation}
This equation can be written explicitly taking into account that the state $|j,j\rangle$ is composed of the states $|m_1,m_2\rangle$ for which the values of $m$'s are restricted by the two natural conditions, $j=m_1+m_2$ and $j-s \leq m_{1,2}\leq s$. The total number of terms $|m_1,m_2\rangle$ that satisfy this condition is $2s+1-j$. Introducing for brevity the notations $k\equiv 2s-j$ and $a_n\equiv C(j,j;m_1=j-s+n,m_2=s-n)$, one can express the state $|j,j\rangle$ as
\begin{equation} |j,j\rangle =\sum_{n=0}^{k} a_n |m_1=j-s+n,m_2=s-n\rangle\, , \label{e3} \end{equation}
Plugging this expansion into Eq.~(\ref{e2}) and taking into account Eq.~(\ref{e1}), one gets:
\[  \sum_{n=0}^{k} a_n \Big\{ [(2s-j-n)(j+n+1)]^{1/2}|j-s+n+1,s-n\rangle \]
\vspace*{-3ex}
\[ +[n(2s-n+1)]^{1/2}|j-s+n+1,s-n\rangle  \Big\} =0 \, . \]

This equation implies the following recurrence relation for $a_n$:
\begin{equation}
a_{n+1}=-r_n a_n \, , \;\; r_n=\Big[\frac{(2s-j-n)(j+n+1)}{(n+1)(2s-n)}\Big]^{1/2} ,  \label{rec} \end{equation}
with $n=0,1, ... , k-1$, and one can check directly that the factors $r_n$ here have the following property:
\begin{equation} r_{k-1-n}=1/r_n \, .   \label{e4} \end{equation}
The recurrence relation with Eq.~(\ref{e4}) make the coefficients $a_n$ in Eq.~(\ref{e3}), roughly speaking, symmetric or antisymmetric with respect to the middle of this sum. More precisely, due to Eq.~(\ref{e4}), the coefficients $a_n$ in Eq.~(\ref{e3}) as determined by the recurrence relation (\ref{rec}):
\[ a_n=(-1)^n a_0\prod_{l=0}^{n-1} r_l \, ,\]
satisfy the condition
\begin{equation} a_{k-n}=(-1)^k a_n \, . \label{e5} \end{equation}

Since the terms $n$ and $k-n$ in the sum (\ref{e3}) are related by the interchange of $m_1$ and $m_2$, we see that Eq.~(\ref{e5}) implies the following symmetry property of the CGCs $C(j,j;m_1,m_2)$, and from this, the CGCs for all other values of $m$:
\begin{equation} C(j,m;m_1,m_2) =(-1)^{2s-j} C(j,m;m_1,m_2)\, . \label{e6} \end{equation}
Finally, combining this equation with Eq.~(\ref{par2}) for the action of the parity transformation, and we see that the magnitude $j$ of the total momentum cancels out from the sign factors, and all the wavefunctions transform in the same way under the particle permutation (\ref{perm}):
\begin{align}
E\psi_{m_1,m_2}(\vec{r}_1,\vec{r}_2) = E_sP \sum_j C(j,m;m_1,m_2) \chi_{j,m} (\vec{r}_1,\vec{r}_2)\nonumber \\
=E_s\sum_j (-1)^{j} C(j,m;m_1,m_2) \chi_{j,m} (\vec{r}_2,\vec{r}_1)  \nonumber \\
= (-1)^{2s} \sum_j C(j,m;m_2,m_1) \chi_{j,m} (\vec{r}_2,\vec{r}_1) \nonumber \\
= (-1)^{2s} \sum_j C(j,m;m_2,m_1) \nonumber \\
\cdot \sum_{\bar{m}_1,\bar{m}_2} C(j,m;\bar{m}_1,\bar{m}_2)\psi_{\bar{m}_1,\bar{m}_2} (\vec{r}_2,\vec{r}_1) \nonumber \\
= (-1)^{2s} \psi_{m_2,m_1}(\vec{r}_2,\vec{r}_1). \;\;\;\;\;\;\;\;\; \label{main} \end{align}
The sums over $\bar{m}_1,\bar{m}_2$ and over $j$ here are taken under the same constraints as in Eqs.~(\ref{CGS1}) and (\ref{CGS2}), and the last step takes into account that the CGSs play the role of both direct and inverse unitary transformations between the states $|m_1,m_2\rangle$ and $|j,m\rangle$. Equation (\ref{main}) establishes the link between the statistics exchange factor of identical particles and their spin not as an extra assumption, but as a direct consequence of the properties of the angular momentum states with respect to the parity transformation in non-relativistic quantum mechanics.

To complete the proof of the non-relativistic spin-statistics connection, one also needs final important step stating, first, that the exchange of identical particles produces the same quantum state of the system by the very notion of the particles being ``identical'' \cite{s01} and, secondly, that the wavefunction should have the same value at the same physical state:
\begin{equation} E\psi_{m_1,m_2}(\vec{r}_1,\vec{r}_2) = \psi_{m_1,m_2}(\vec{r}_1,\vec{r}_2)  \, .\label{last} \end{equation}
Combined with Eq.~(\ref{main}), this equation gives the usual form of the spin-statistics relation:
\begin{equation} \psi_{m_1,m_2}(\vec{r}_1,\vec{r}_2)=(-1)^{2s} \psi_{m_2,m_1}(\vec{r}_2,\vec{r}_1). \end{equation}

Equation (\ref{last}) is based on the requirement on the wavefunction of a quantum system to be single-valued. This requirement underlies all attempted demonstrations of nonrelativistic spin-statistics relation, and was debated most clearly in the case of the simplest instance of this relation: spinless bosons. Configuration space of point particles without spin in 3 spatial dimensions is defined by their spatial coordinates only, and in this case, the condition that the wavefunction is single-valued is the only necessary element of the proof of the spin-statistics conncection. Permutation of identical particles in 3 spatial dimensions clearly produces an identical state, so that the single-valued nature of the wavefunction immediately implies the same value of the wavefunction upon permutation, and hence,  bosonic exchange statistics \cite{s02}. Although the requirement for the wavefunctions to be single-valued can be viewed as controversial \cite{s03}, it represents the only way to account for many essential features of quantum mechanics, and should be accepted as one of its fundamental postulates. Frequent emphasis on the fact that quantum states are defined only up to an overall phase is valid only for isolated systems, and looses its validity as soon as the system is made to interact with another quantum system. All the current developments in quantum information science emphasize the fact that the phase of a quantum state is a well-defined physical quantity, and the states that differ by an overall phase are by no means identical. In view of all this, the fact that the permutation of identical particles produces precisely the same vector in a Hilbert space should be viewed as a fundamental principle of quantum mechanics.

Finally, it is useful to illustrate the arguments presented above for arbitrary spin $s$ by two simple and most important examples of small spins s=1/2 and s=1. As is well known, the two spin-$1/2$ particles can form a triplet state with total momentum $j=1$, which is symmetric with respect to interchange of the spin variables, and a singlet state with $j=0$, antisymmetric in spins. The parity transformation in the center-of-mass coordinate system interchanges the spatial coordinates of the two particles and multiplies their total spin state by the usual factor $(-1)^j$, i.e. 1 for the singlet, and -1 -- for the triplet states. This ensures the fermionic exchange statistics as a consequence of the properties of the angular momentum.

One can also check explicitly the CGSs for addition of the two spins $s=1$ (in the same notations as used above)
\begin{align} C(22,11) =1, \;\; C(21,10)=C(21,01)=1/\sqrt{2}, \nonumber  \\
C(20,-1,1)=C(20,1,-1)=\frac{1}{\sqrt{6}}, \; C(20,00)=\sqrt{\frac{2}{3}}, \nonumber \\
C(00,-1,1)=C(00,1,-1)=\frac{1}{\sqrt{3}}, \; C(00,00)=-\frac{1}{\sqrt{3}} , \nonumber \\
C(11,1,0)=-C(11,0,1)=1/\sqrt{2}, \nonumber \\ C(10,1,-1)=-C(10,-1,1)=1/\sqrt{2} .  \nonumber
\label{ex} \end{align}
to see that the even-$j$ coefficients are even, while the odd-$j$ are odd with respect to the interchange of the two spins.
Combined with the similar property of the parity exchange of the orbital coordinates, this ensures the bosonic exchange statistics of spin-1 particles, again as a consequence of the properties of the angular momentum.

In conclusion, this work suggests the derivation of the spin-statistics relation for identical particles with an arbitrary spin $s$ in 3 spatial dimensions directly from the properties of the angular momentum in non-relativistic quantum mechanics. The central role in the suggested arguments is played by implementation of the permutation of the spatial coordinates of the two particles through the parity transformation in the center-of-mass coordinate system. This approach can be directly extended to the systems of more than two identical particles by consecutively considering all pairs of particles.

This work was supported by the US NSF grant \# 2104781.

\end{document}